\documentclass[12pt]{article}
\usepackage{mathtext}
\usepackage{amsfonts}
\usepackage{amssymb}
\usepackage{amsmath}
\renewcommand{\d}{\delta}

\newcommand{\g}{\gamma}

\newcommand{\e}{\epsilon}

\newcommand{\ar}{\longrightarrow}

\newcommand{\la}{\lambda}
\renewcommand{\a}{\alpha}
\begin{document}
\title{Dynamic diffusion as approximation of quantum behavior}
\date{}
\author{Y.I.Ozhigov\\
\\
{\it Moscow State University of M.V.Lomonosov}\\
{\it Institute of Physics and Technology RAS}}
\maketitle

\begin{abstract}
The approximation of quantum unitary dynamics of a particle by a swarm of point wise classical samples of this particle is proposed. Quantum mechanism of speedup rests on the creation and annihilation of absolutely rigid bons, which join samples in dot wise symplexes so that the density of swarm approximate the quantum probability. This mechanism does not require differentiation of a density that is adventage of this method over Bohm's quantum hydrodynamics: our method is applicable to many particles in entangled states. In multi particle case the limitation of total number of samples gives the natural model of decoherence, e.g. the divergency from the exact solution of Shredinger equation. Intensity of creation - annihilation of bonds between samples substantially depends on the grain of spatial resolution, which makes impossible to pass to the limits as in a classical substance; this is the price for the scalability of a model to many particles.  
\end{abstract}

\section{Introduction}

Interpretations of quantum mechanics will be actual still many years because they are aimed to the creation of models of complex objects, which behavior has quantum roots - up to living things. However, the standard quantum mechanics yet is not ready for it. The most diffucult obstacle is that the dimensionality of the space of quantum states grows exponentially when the total number of particles grows. For that aim, accordingly to the proposal of R.Feynman and others (see \cite{Fe}) we should build the so called quantum computer. But the practical experiments with limited quantum processors show that the physics of a quantum computer itself represents the principle problem, and the known contra intuitive character of Copengagen quantum theory does not permit us to penetrate into this new area. 

Quantum physics of many particles thus represents the new challenge, which requires to modify quantum theory along the direction to the common sense. This modification must include the notion of trajectory, existence of dynamic characteristics as velocity and coordinates at the same time, and the explicit, ''urn'' scheme for quantum probability where the probability to find a particle in some state could be computed as the fraction $N_{suc}/N_{tot}$ accordingly to the frequency definition of a probability. A model with all these features can be called an interpretation of quantum mechanics (in semi classical terms). The way to the classical interpretation of quantum mechanics lies through Feynman path integrals, where a quantum state in each time instant is represented as a swarm of point wise particles, each of which $\a$ besides coordinates and velocity carries also the phase $\phi_{\a }$. At each step the wave function $\Psi=|\Psi |e^{i\phi_s}$ is divided to small portions of the form $\e e^{i\phi_s}$ and $|\Psi |/\e$ point wise samples are introduced in the corresponding cell of space, each of which obtains the phase $\phi_s$ and velocity distributed uniformly over all space. The all samples fly with the corresponding speeds and change their phases by the addition of $\frac{\Delta S}{h}$ where $\Delta S$ is the change of the action along the way they passed at this step, $h$ is Plank constant. Then we sum up the amplitudes $\e \ exp(i\phi)$ of all samples occurring at the same cell that gives the new wave function, then repeat the step again, etc. Here samples live only during one step, at the same step we define them anew. 

Feynman path integrals in this discrete formulation makes possible to speak about trajectories and to compare the different trajectories; for example it gives the border of applicability of classical mechanics: if the action along the classical trajectory exceeds Plank constant we should use classical description, if the action becomes comparable with $h$ we should use quantum mechanics. Feynman path integrals is the approach completely equivalent to quantum mechanics, though it is not its interpretation in our sense: it does not give us the ''urn'' scheme for quantum probability. The deep reason of it is the abcense of dynamical characteristics of samples: their speeds has the marginal distribution and all information about the dynamics is contained in their phases and their amplitudes, which are in addition the subject for summing that eliminates the ''memory'' of former samples at each step. 

We thus conclude that the main feature of samples, which path integrals do not have, is the conservation of their memory. Samples must have their own histories, separate from each other. Only with this property would be possible to obtain the interpretation of quantum mechanics we need. 

\section{Bohm approach}

The problem of conservation of samples individuality has been solved by D. Bohm, who proposed the swarm of classical particles, which are driven by the external potential $V=V(\bar r)$ and the so called quantum potential $Q=-\frac{h^2}{2m}\frac{\Delta R}{R}$, where $\Psi = R\ exp(iS/h)$. Shredinger equation with the external potential $V$ is equal to two real equations of the form:
\begin{equation}
\begin{array}{ll}
\frac{\partial S}{\partial t}+\frac{p^2}{2m}+V+Q&=0,\\
\frac{\partial \rho}{\partial t}+\nabla\cdot (\frac{1}{m}\rho p)&=0,
\end{array}
\label{HJ}
\end{equation}
where the first is Hamilton-Jacoby equation for the characteristic function $S$ of a sample, which is defined by the equations $\nabla S=p, \ \partial S/\partial t=E$, where $p=p(\bar r)$ is its impulse, $\rho (\bar r)= R^2 (\bar r)$ - the density of swarm, and the second equation shows the property of continuosity of a swarm. This approach is called quantum hydrodynamics, because a quantum particle is here represented by a swarm of its classical samples, which obeys the law of classical dynamics but their potential $U=V+Q$ has the quantum summand $Q$, which depends on the density of the swarm in this point in the time and space. 

The algorithm modelling quantum dynamics by hydrodynamical approach at each step does the following. 
\begin{itemize}
\item Fulfills the free flight of all samples.
\item Change the speed of each sample accordingly to the rule $p(t+\Delta t)=p(t)-\nabla U$.
\end{itemize}

We see that the computation of the density of swarm at each step is very important, because it is required for the computation of $U$, due to the addition in the form of quantum pseudo potential $Q$. We can consider that samples merely determin the nodes of division of configuration space, which is the base of the finite differences method for the solution of (\ref{HJ}); the value of quantum hydrodynamics is thus limited by the possible acceleration of the solution of Shredinger equation. How valuable is this adventange of quantum hydrodynamics? We consider the case of many real particles, for example a few tens of them. The density of swarm will be then the density of swarm in $R^{3n}$ space, where $n$ is the total number of particles. To conserve the initial accuracy of the approximation to 

\begin{equation}
\rho (t)\approx |\Psi (t)|^2
\label{rho}
\end{equation}
 for the solution $\Psi$ of Shredinger equation, we need to keep constant linear spatial step, which results in the exponential grows of the total number of cells in the configuration space. To be able to compute derivatives from the density (we need three sequential) we must guarantee that each cell is full of samples that erects the same difficulty as the direct solution of Shredinger equation for $n$ particles. The single profit from the quantum hydrodynamics consists in the possibility to optimize the grid for the finite differences scheme; this adventage is interesting but it does not resolve our main problem. 

\section{Swarm with dynamic diffusion}

Explicit consideration of mechanism for the acceleration of samples requires the satisfaction of conservation laws. If we are going to refuse from the knowledge of density in the vicinity of the point at hand, we must ensure these laws for separate samples, which means the refusal from such things as the quantum pseudo potential. First of all we define this mechanism and write the equation on the string of samples through the border, which this mechanism induces; then we show, that Shredinger equation can be reduced to this equation if we fix the grain of the spatial resolution. It turns that our mechanism of dynamic diffusion critically depends on the choise of this grain that causes some peculiarities of our approach. However, these peculiarities are inavoidable price for adventages we obtain with dynamic diffusion for many particles.  

\subsection{Mechanism of bond creation and annihilation}

We will define the evolution of swarm, which we call the dynamic diffusion and which will consist of steps, each of duration $\Delta t$. But here the simple scheme from the previous section is not sufficient, because the swarm will substantially transformed, hence we define the step of evolution anew.

At each step we will have not samples, but symplexes. A symplex $S$ of an order $j$ ($j\in N$) is a point wise particle, which has its own coordinates $\bar r(S)$, speed $v(S)$, and time of waiting $\tau (S)$. Simultaneously, we will define at each step the new objects called bonds. Speaking non formally, a symplex is some set of samples connected by bonds with each other, which roll with very hight speed so that their kinetic energy is large whereas the speed of the symplex as the whole is low. Factually, the swarm will be divided to two fractions: separate samples, travelling with the largest possible speed $c$, and symplexes of orders $j>1$, which speed is much lower. This is the reason of the itroduction of the time of waiting: this is the time, during which the symplex $S$ flies, but which is not enough that $S$ can overcome the distance to the neigboring cell. As we will see in the next section the fixation of the division of space to cells is necessary for the simulation of quantum dynamics; this is why we cannot consider the continuous movement of symplexes. 

For this we first of all define the procedure of association of a sample $s$ with a symplex $S$ of the order $j$. Let $S$ be located in the cell $\bar r$, and $s$ be a sample, which has just flied to this cell. We suppose that during the considered step the distance between $s$ and $S_j$ reaches its minimal value. We now suppose that the time flies continuously, it will not change the duration of step, because of the great fraction $c/v(S)$. At the instant when the minimal distance is reached we establish the absolutely rigid connection between $s$ and $S$ so these objects start to roll classically around their common center of masses, the resulted object we call a symplex of the order $j+1$. Its speed $v_1=(v(s)m+v(S)M)/(m+M)$ where $m,\ M$ are masses of $s$ and $S$ correspondingly, results from the classical consideration of the system $s+S$, the single non classical object here is the bond we introduced between $s$ and $S$. We call this bond the main bond of the resulted symplex $S_1$ and designate it as $b(S_1)$. We note that due to the absolute rigidity of the bond it does not produce any work; the energy, impulse and momentum of impulse are thus conserved in course of assotiation. 

The process of bond elimination is exactly reverse of the association; it results in the separate sample, which flies away from the symplex that becomes less to one in the order. The iteration of the associations $q$ times starting with $S$ gives us the symplex $S'$, which order is $j+q$. Factually, a symplex of the order $j$ represents the sequentially nested symplexes, each of which is the result of association between some sample and the predecessing symplex. The sequential dissociation goes in the reverse order. 

Now we are ready to define the step of the main evolution. 
The step looks as the following sequence. 
\begin{itemize}
\item For each symplex $S$ we change $\tau (S)$ to $\tau (S)+\Delta t$ and check the inequality $\tau (S)v(S)<d$ where $d$ is the distance to the closest neigboring cell along the direction of $v{S}$. If it is satisfied, we go to the next action, if it is violated, we shift $S$ to the closest neigboring cell and put its time of waiting to zero: $\tau (S)=0$.
\item Each symplex obtains the additional speed as $b\d t \nabla V/m$, where $V$ is some external potential, $\d t$ is the duration of step, $m$ is the mass of a sample, $b$ is some constant.
\item From each symplex $S$ we take $[aj]$ last samples from $S$ and sequentially eliminate bonds connecting them with the symplex, call them belonging to thin layer and let them fly to the closest neigboring cells with the speeds they have, where $j$ is the order of this symplex, $a$ is some constant $a\ll 1$. 
\item After the arrival of each sample belonging to thin layer to the closest neigboring cell we associate it with the corresponding symplex and recalculate its speed by the law of association.
\end{itemize}

We call the effect of bonds annihilation by peak explosion, because it seems as the explosion of the symplex, where the hidden energy of rolling samples transforms to the visible kinetic energy of released samples, which fly out of the cell at hand. The third item says that these samples are immediately absorbed by the neigboring symplexes; their energy and impulse is thus captivated by these symplexes. Peak explosion bears the resemblance with the gas fraction, the movement of symplexes following from the first point we can call the liquid stream. The first and second items establish the rules for phase transition: from liquid phase to the gas and vice versa. The velocity of gas fraction is much higher than of the liquid phase; this is the representation of non relativistic character of qynamics we regard:
\begin{equation}
v_{symp}\ll c.
\label{nonrel}
\end{equation}
It causes the principal difficulty when $\rho$ becomes too small, because here the corresponding symplex will obtain the speed comparable with $c$. In this case our definition of symplexes loses its validity at all because this definition presumes that flying samples appears to a cell when a symplex stays; this symples cannot leave the cell until the sample associates with it. We note that this difficulty is the same, as in Bohm hydrodymanical approach; it is hardly possible to overcome it with swarms. On the other hand, the area where $\rho =0$ plays the principal role in the simulation of quantum dynamics; its influence shows immediately because of a great speed of gas fraction. In a quantum swarm, no difference how it is treated (hydrodynamics or dynamic diffusion) it is impossible to make consideartions local, as in ordinar hydrodynamics when we separate ''small cube'' and consider a stream through its border. The reason is the existence of a gas fraction. Further we will have the more quantitative characteristic of this form of quantum non locality.

We choose the initial state of the swarm in such a form, where each cell is occupied by exactly one symplex; the density $\rho(\bar r)$ will be then proportional to the order of the symplex occupying the cell $\bar r$. 

The evoltion we have defined depends on the constant $a$, which is the intensity of bonds annihilation; we can treat it as the $1/\Delta T$ where $\Delta T$ is the average time of bonds life. The process of bond annihilation will be thus the result of Poisson random process with the given intensity. 

We now consider two neigboring cells $x$ and $x_1$ with the common border, the density of stream through this border $p_{x,x_1}$, and find its variation in the time. Density of stream is the stream divided to the square $\d x^2$ of the border (that is the stream through the border of a unit sized cube - we take it to pass from quantity of samples to the density), we call it simply the stream. Why we focus at the stream variation, not the stream itself? Because the stream depends on the initial state of the swarm, which must be given in advance: we cannot derive it from the mechanism of samples speedup. The mechanism of dynamic diffusion we defined influences to the stream only through its variation, which is the dynamical magnitude, as a force. We will see, that in the initial moment the stream is created by the movement of symplexes, whereas separate samples belonging to thin layer create the variation of stream. 

Let $j$ and $j_1$ be the orders of the symplexes in the cells $x$, and $x_1$ correspondingly. 
We must estimate the deposits to the steam variation from gas and liquid fractions separately. 

1). {\bf Liquid fraction - inertia}. We estimate the deposit of the first item. The function of density $\rho$ typically has the strongly discontinuous form, where ''peaks'' are interspered with ''holes''. Of course, it is the influence of the fixation of $\d x$; further we recognize that it is inavoidable. To cope with this discontinuity we can reform the first item, and let samples fly by the annihilation of bonds, as in the second item. Here the part of $a_lj$ samples will fly to the direction $\bar l$ where $a_l=a_0\ \bar l\cdot v(r)$, where $a_0$ is small constant, $a_0\ll a$. This way of making $\rho$ more continuous, however, contradicts to the invariance towards the change of inertial readout system: our mechnism must not depend on this choice. We make this agreement temporarily, in order to estimate the deposit of the liquid phase to the change of stream. Thus, this deposit will be $v_{norm}(r)\nabla\rho (r)$, where $v_{norm}$ is the component of speed of the swarm, orthogonal to the border, because it comes from the variation of the order of symplexes, which comes through this border. 

2) {\bf Liquid fraction - external potential}. This deposit equals $b\rho\nabla V$. 

2) Gas fraction. The deposit of peaks explosion is  the In view of the second item of our definition of a step, this stream equals the difference between the number of samples flied from the cell $x$ to $x_1$, and the number of samples flied in the opposite direction, that is the difference between $j$ and $j_1$. The deposit of gas fraction is thus $a\nabla\rho$. 

We can make the following conclusion. If constants $a $ and $b$ are substantially larger than $a_0$, the change of the stream in the unit of time is following equation
\begin{equation}
\frac{d p_{x,x_1}}{d t}= a\nabla \rho + b\rho\nabla V.
\label{detailed_stream}
\end{equation}

This equation characterizes dynamic diffusion. To show that this mechanism can serve as the approximation of quantum unitary dynamics, we have now to derive it from Shredinger equation.

\subsection{Reducing of Shredinger equation to swarm}

We now consider the swarm, for which (\ref{rho}) is true where $\Psi$ is exact solution of Shredinger equation, and try to derive the evaluation of the stream variation from Shredinger equation directly, accepting some features of sample moving, which agree with the mechanism of dynamic diffusion. Our aim is to show that this variation satisfies (\ref{detailed_stream}). 

To ensure the main requirement (\ref{rho}) we must define swarm density as
\begin{equation}
\rho (r) = \frac{N_d}{dx^3N_{total}},
\label{rho_def}
\end{equation}
where $N_d$ is the number of samples occurring in the cube with the side $dx$ with the center $r$, $N_{total}$ is the total number of samples in the swarm.
To compare with the solution of Shredinger equation we would have to launch in this definition $\d x \ar 0$, which would mean that we consider not one swarm but the sequence of swarms with densities $\rho_n$ with increasing $n$. We will not do it in order to avoid the useless complification, instead each time when it is necessary we agree that it is possible to continue our division of the space to cubes so that $\d x$ will decrease in the admissible limits. We write $\rho(x)=|\Psi(x)|^2$, which means that  
\begin{equation}
\rho_n(x)\ar |\Psi(x)|^2 \ \ (n\ar \infty ),
\label{asympt}
\end{equation}
without special mentioning. Such a sequence of swarms realizing the approximation to the density of wave function - solution of Shredinger equation we call the admissible approximation of quantum evolution. Now having this agreement, we fix the grain $\d x$ of spatial resolution. This fixation is necessary for the computing of stream variation. We call the stream with the fixed $\d x$ the detailed stream, to emphasize its substantial dependence from $\d x$. 

At first we show that there exists the quantum swarm with the local shifts of samples, e.g., that the equality (\ref{rho}) can be guaranteed by only shifting samples on the close distances: between neigboring cells.

We would easily guarantee the satisfaction of (\ref{rho}) if we do not impose any limitations on the speeds of samples and on its change. It means that samples can move in course of one step on any distance and the equation (\ref{rho}) will be true on each step with the required accuracy; it will be conserved in some time frame $\d t$. However, this mechanism is useless because it depends on the knowledge of the wave function whereas our aim is to manage without the wave function at all.

We take up the quantum swarm, and start from Shredinger equation
\begin{equation}
ih\frac{\partial\Psi(r,t)}{\partial t}=-\frac{h^2}{2m}\Delta\Psi(r,t)+V_{pot}(r,t)\Psi(r,t),
\label{Sh}
\end{equation}
which we can rewrite as  
\begin{equation}
\begin{array}{lll}
&\Psi^r_t(r)&=-\frac{h}{2m}\Delta\Psi^i_t(r)+\frac{V_{pot}}{h}\Psi^i(r),\\
&\Psi^i_t(r)&=\frac{h}{2m}\Delta\Psi^r_t(r)-\frac{V_{pot}}{h}\Psi^r(r)
\end{array}
\label{Sh2}
\end{equation}
for the real and imaginary parts $\Psi^r,\ \Psi^i$ of the wave function $\Psi=\Psi^r+i\Psi^i$. We are interested in the evolution of density only, e.g., the function 
$$
\rho(r,t)=(\Psi^r(r,t))^2+(\Psi^i(r,t))^2.
$$

We fix the value $\d x$ and apply for the approximation of second derivatives the difference scheme of the form
$$
\frac{\partial^2 \Psi(x)}{\partial x^2}\approx\frac{\Psi(x+\d x)+\Psi(x-\d x)-2\Psi(x)  }{(\d x)^2}
$$
for each time instant, provided the wave function satisfies all conditions for such approximation. Since the addition of any constant to the potential energy $V_{pot}$ does not influence to the quantum evolution of the density, we can consider instead of $V_{pot}$ the other, equivalent potential  $V_{pot}+\a$, where $\a=-\frac{3h^2}{m(\d x)^2}$ that results in the dissapearence of the summand $2\Psi(x)$ in the difference schemes for second derivatives on $x,y,z$ (from that the coefficient $3$ appears) after its substitution in Shredinger equation.\footnote{This trick is not accurate from the mathematical view point: we use the fact (possibility to add a constant to potential), which is substantiated by analysis, whereas we fix $\d x$ and in further cannot launch it to zero. However, it is not critical: we could preserve the last summand and fulfil computations with it; it gives the same result but complicates computations.} For the simplicity of notation we introduce the coefficient 
$$
\g = \frac{h}{2m}\frac{1}{(\d x)^2}.
$$
Since we yet do not know the mechanism of moving of samples in the quantum swarm, we suppose that we simply take off some quantity of samples from one cell or place them to this cell from some storage. We divide the evolution of quantum swarm to so small frames of the longitude $\d t$, that on each frame samples travel in the framework of two neigboring cells. If we prove that the diffusion mechanism provides the evolution on each of such frames, it will be true for the whole evolution because our supposition about the exchange between two closest cells do not limit the generality. We also agree that these cells differ from each other on the shift to $\d x$ along the axis $x$, which does not limit the generality as well. We denote centers of these cells by $x$ and $x_1=x+\d x$. Due to our suppositions about the exchange the summand $\Psi(x-\d x)$ in the difference scheme disappears as well and on the short time frame the evolution is determined by the following system of equations:
\begin{equation}
\begin{array}{lll}
&\Psi^r_t(x)&=-\g\Psi^i(x_1)+V(x)\Psi^i(x),\\
&\Psi^i_t(x)&=\g\Psi^r(x_1)-V(x)\Psi^r(x),
\end{array}
\label{quant}
\end{equation}
and the analogous system obtained by the replacement of $x$ by $x_1$ and vice versa, where the lower index means the differentiation on $t$. 
It means that we regard only the exchange of samples through one fixed border between cells $x$ and $x_1$. If we then sum the other deposit corresponding to the exchange between $x$ and $x_0=x-\d x$, we obtain the full change of density within the effect of the separation to two fractions (see below). This exchange has the analogous form and all further computations can be fulfilled for it with the corresponding result; we thus take up only exchange between $x$ and $x_1$.

The system (\ref{quant}) is true in the supposition that samples move from $x$ to $x_1$. When samples move from $x_1$ to $x$ we obtain the analogous system obtained by the replacement of $x$ by $x_1$ and vice versa. Shredinger equation will then express the result of the general evolution process consisting of the both cases where $x$ and $x_1$ can settle down by six ways along three coordinate axis. By the time frame $\Delta t$ we now mean just such short time segment when the exchange goes only between $x$ and $x_1$ (and, may be, the storage; we will shortly see that the storage is not needed).

For such a segment we have
\begin{equation}
\begin{array}{lll}
 p_{x,x_1}= \frac{\partial\rho(x)}{\partial t}|_{x,x_1}&=2\Psi^i(x)(\g\Psi^r(x_1)-V(x)\Psi^r(x))&+2\Psi^r(x)(-\g\Psi^i(x_1)-V(x)\Psi^i(x))=\\
&=2\g (\Psi^i(x)\Psi^r(x_1)-\Psi^r(x)\Psi^i(x_1))&=-\frac{\partial\rho(x_1)}{\partial t}|_{x,x_1},
\end{array}
\label{1der}
\end{equation}
where by $\frac{\partial\rho(x)}{\partial t}|_{x,x_1}$ we denote the deposit to the derivative of density on the time, which appears from the moving of samples through the border separating cubes that contain points $x$ and $x_1$. 
 It gives that the decrease of samples in one cell equals the increase of them in the other, e.g., the evolution of quantum swarm satisfies the condition of locality and we can speak about the stream of samples through the element of surface when we fix the value of $\d x$.

\subsection{Dependence of swarm dynamics from grain}

We approach to the most important thing in the description of quantum dynamics: the dependence on the grain $\d x$. We cannot confirm that the expression (\ref{1der}) determines the number of samples, which increases the density of the corresponding cell when they pass through the considered surface in the unit of time. Expression for a detailed stream (\ref {1der}) is not expression for an ordinary stream of particles through a surface. The reason in that it critically depends on a value $ \d x $, and it is impossible to direct this value to zero. We cannot apply to a swarm the fundamental reception of the mathematical analysis, consisting that it is possible to divide unlimitly intervals $ (x, x_1) $ and to pass to a limit at $ \d x\ar 0 $ so all values of group magnitudes (the density and speed) will keep the values (and even them will specify). Expression (\ref {1der}) indirectly specifies that at the swarm there are samples with essentially different speeds so we cannot speak about speed of all swarm in the given point. Let's recollect that the swarm has two parts: fast (separate samples from the thin layer) and slow (symplexes). The mechanism of moving of samples of the fast part consists that each of them on each step $ \Delta t $ jumps in a direction of its speed on the number of cells proportional $ \Delta t $. The mechanism of moving of a sample of the slow part of the swarm consists that it waits number of the steps, proportional 1$/\Delta t $, being at a stop, and only then moves exactly on one step in a direction of the speed. In this case we cannot write for the detailed vector stream (density) $\bar p$ expression 
\begin{equation}
\frac{\partial\rho(r,t)}{\partial t} =\frac{1}{N (\d x)^3} \int\limits_{S(r)}\bar p(r,t)\bar n(\bar r_1)ds(r_1)
\label{us_st}
\end{equation}
that is true for the ordinar stream $p$. Instead we must use the expression
\begin{equation}
\frac{\partial\rho(r,t)}{\partial t} =\frac{1}{N (\d x)^3} \int\limits_{S(r)}\bar p(r,t)|_{(x,x_1)}\bar n(\bar r_1)ds(r_1)+\Lambda,
\label{det_st}
\end{equation}
where the summand $\Lambda$ have the following nature. 

Let $\d x$ and $\Delta t$ be fixed. To write expression (\ref {us_st}), it is necessary to guarantee that all samples of the fast part of swarm, which have past  through the border in a direction $r $, remain in the corresponding cell, and do not jump out of it because of the great speed. For this, it is necessary to make $ \Delta t $ small enough. But then samples of the slow part of the swarm will not have time to move a little at all! For maintenance of movement both fast, and slow samples there is only one way: to reduce a grain $ \d x $. But we also cannot make it, because then at once will increase speeds of the fast part of the swarm, and we should consider already the other swarm.

Thus, it is impossible to replace discrete character of moving of samples of fast and slow parts of the swarm with continuous movement, as in case of the classical continuous environment. Summand $ \Lambda $ it makes sense the phase transition between fast and slow parts of the swarm. We turn at the necessity to consider two fractions: slow liquid and fast gas. (Otherwise, if we make possible for symplexes move at each step, we would have to consider jumps of a thin layer samples: the jump of a fast sample means that it disappears in one cell then appears in other cell which can have no common borders with the first.) There is no the usual stream in a quantum swarm, we can thus operate with the detailed stream only. 

Inapplicability of usual analytical receptions to the quantum swarm does not mean impossibility of a substantiation of a way of modelling offered by us by means of classical mathematics at all.\footnote {In a case the delta of function of Dirac the classical mathematical substantiation formally has been found in the form of linear functionals.} I do not exclude a finding of such substantiation, but - only for the case of one real particle. However, it is impossible to find a classical substantiation of the method of collective behaviour in its general form, for many particles, because here the limitation of quantity of samples plays the central role that makes  impossible application of the ideology of the mathematical analysis. 
It, in particular, means also hopelessness of attempts to prove formally that our mechanism of moving of samples of the swarm in all cases gives good approximation of the exact solution of Shredinger equation. Therefore the following below substantiation of the proposed mechanism should be treated as the explanatory to algorithm work, but not as the strict proof of its suitability in all cases. The general ideology of constructivism (see \cite{Oz}) offers us one: to rely on direct computer modelling.

\subsection{Reduction of Shredinger equation to dynamic diffusion}

We introduce the following parameters depending on $\d x$:
\begin{equation}
I=\frac{h^2}{2m^2 (\d x)^3},\ \kappa =\frac{h}{m\ \d x}
\label{intens}
\end{equation}
where $h$ is Plank constant. The parameter $I$ is called the intensity of action of the density gradient, $\kappa$ is the intensity of potential. 

Now for construction of the mechanism of change of speed of samples we will find the change of the detailed stream
$\frac{\partial}{\partial t}p|_{(x,x_1)}\ \bar ndS $ of quantum swarm through a surface of a small cube, the normal to which is parallel to the axis $OX$. For this purpose it is necessary to differentiate the expression (\ref{1der}) on the time:
\begin{equation}
\begin{array}{ll}
&\bar p'_t=2\g [ (\g\Psi^r(x_1)-V(x)\Psi^r(x))\Psi^r(x_1)+\Psi^i(x)(-\g\Psi^i(x)+V(x_1)\Psi^i(x_1))-\\
&(-\g\Psi^i(x_1)+V(x)\Psi^i(x))\Psi^i(x_1)-\Psi^r(x)(\g\Psi^r(x)-V(x_1)\Psi^r(x_1))]=\\
&2\g^2(\Psi^r(x_1))^2-2\g V(x)\Psi^r(x)\Psi^r(x_1)-2\g^2(\Psi^i(x))^2+\\
&2\g V(x_1)\Psi^i(x)\Psi^i(x_1)+2\g^2(\Psi^i(x_1))^2-2\g V(x)\Psi^i(x)\Psi^i(x_1)-\\
&2\g^2(\Psi^r(x))^2+2\g V(x_1)\Psi^r(x)\Psi^r(x_1)= \\
&2\g^2((\Psi^r(x_1))^2+(\Psi^i(x_1))^2-((\Psi^r(x))^2+(\Psi^i(x))^2))+\\
&2\g [(V(x_1)-V(x))((\Psi^r(x))^2+(\Psi^i(x))^2)+o(\d x)],
\end{array}
\end{equation}
where $o(\d x) = (\Psi^r(x)\Psi^r(x_1)+\Psi^i(x)\Psi^i(x_1)-((\Psi^r(x))^2+(\Psi^i(x))^2))(V(x_1)-V(x))$. 

We thus can write for the detailed stream the formula 
\begin{equation}
\Delta p|_{(x,x_1)}=(-I\nabla\rho -\kappa\rho\nabla V)\Delta t.
\label{stream_detailed}
\end{equation}

We see that Shredinger equation gives the same change of stream as the dynamic diffusion (see equation (\ref{detailed_stream})) if we put $a=-I,\ b=-\kappa$. 

\subsection{Restoration of wave function from dynamical diffusion swarm}

For a conformity finding between the standard description of a quantum state through its wave function $ \Psi $ and its swarm representation, we will assume that the last is described by pair of functions
\begin{equation}
\rho(t,\bar r),\ \bar p(t,\bar r),
\label{pair}
\end{equation}
where the scalar function $\rho$ is the density of samples, and the vector of impulse of the swarm $\bar p(r)$ is the sum of speeds of samples occurred in the small cell with the side $\d x$ and the center in the point $r$. The swarm impulse does not include weight of the real particle; $p (r) $ there is the swarm characteristic in which ''the weight'' role plays the total number of samples.

Such pair does not use concept of complex number, and does not give the beautiful differential equations of Shredinger type for $ \rho $ and $ \bar p $. Moreover, the mechanism of dynamic diffusion introduced by us for imitation of quantum evolution, considerably differs from classical processes (for example, a heat transfer or fluctuations) that its intensity depends on the chosen grain of the spatial resolution. We have gone on it for the sake of the main thing: economy of computing resources which are not simply necessary for modelling of quantum dynamics of complex systems, but are absolutely necessary for its research in general.

Given a state of swarm we will show, how the usual complex wave function $ \Psi $ can be restored from the pair (\ref {pair}). We accept that the carrier of wave function is connected, that is any two points in it can be connected by a curve which is not crossing area of zero density $ \rho=0$. This restriction is related to that is available in the diffusion Monte Karlo method.
For this purpose we consider the equality (\ref {1der}), and substitute in it the expression for wave function through density: $ \Psi (r) = \sqrt {\rho (r)} \exp (i\phi (r)) $. The problem consists in the calculation of the phase $ \phi (r) $ of wave function. We will notice that as the relative phase between various points has physical sense only, we can fix some point $r $ and consider a phase of other point $r_1$ relatively to $r $. If $r_1$ is close to $r $, the equation (\ref {1der}) gives us
$$
\phi(r)-\phi(r_1)=arcsin\ k(\d x)^2\frac{\bar p(\bar r-\bar r_1)}{\sqrt{\rho(r)\rho(r_1)}}
$$
that leads to the following formula for the relative phase:
\begin{equation}
\phi(r_1)=\int\limits_\g k(\d x)^2\bar v\ d\bar\g
\label{phase}
\end{equation}
where the path $\g$ goes from $r$ to $r_1$. This equation explicitly depends on the choice of the path $\g$ hence we have to prove its correctness, e.g., independency from the choice of $\g$. 

We mention that this derivation will be correct only in the case when $\rho > e>0$ for some constant $e>0$, e.g., the density must be separated from zero in all area of consideration. Since the phase is determined only within to the integer multiple of $2\pi$, different choices of the path can lead at most to the choice of the phase to such a number that takes place, for example, for excited states of an electron in hydrogen atom with nonzero magnetic number. We show that the integration of the speed $\bar v$ of swarm along a closed path preserves its value in the time the more exactly, the lesser grain of spatial resolution $\d x$ is. It results that if in the initial time instant the definition (\ref{phase}) is correct it preserves the correctness for the following instants as well.

We thus consider the derivative of the integral of the speed of swarm along the closed path $\g_c$. Applying the formula (\ref{stream_detailed}) and taking into account that $\partial\bar p /\partial t$ is proportional to $\rho\ \partial\bar v/\partial t$, we obtain 
\begin{equation}
\frac{\partial}{\partial t}\int\limits_{\g_c}\bar v\ d\g=-\int\limits_{\g_c}A\ I(\d x)^2\frac{\nabla \rho}{\rho}+B\kappa (\d x)^2\ \nabla V
\end{equation}
for some $A,\ B$. 
The first summand gives zero after the integration along the closed path because it is $\nabla \ln\rho$, the second summand gives zero by the analogous reason. 

It is now sufficient to check that the definition (\ref{phase}) is correct in the initial instant that can be done straightforwardly for each task. In the case when the wave function of initial state can be obtained from the ground state of electron in hydrogen atom, where $\bar v=0$, the correctness follows from the proven because there is no phase shift to $2\pi k$. The phase of each groundstate does not depend on the point. If for the obtaining of the initial state in the considered problem we have to start from some escited state we must at first check the correctness for this state. 

On the basement of these computations we can write formulas connecting swarm parameters with the wave function:
\begin{equation}
\begin{array}{ll}
&|\Psi(r)|=\sqrt{\rho(r)};\\
&\phi(r)=\int\limits_{\g :\ r_0\ar r}k\bar v\cdot d\g,\\
&\bar v = b\nabla\phi(r),
\end{array}
\label{connection}
\end{equation}
for some $a,\ b$.
These formulas permits to pass from the wave function to the swarm and vice versa. This description has two features. At first, beyond Shredinger dynamics on the level of grain $\d x$ stays the swarm dynamics of the lower level with grains $\Delta x,\ \Delta t$, so that there is the substantial dependence of swarm parameters (intensity) from the chosen grain of spatial resolution $\d x$. At second, quantum dynamics presupposed the presence of the thin layer and the specific behavior of samples towards it. 

In formulas (\ref{connection}) the mean speed $\bar v$ of swarm approximately equals the mean speed of such a part of swarm, which does not contain thin layer in each point where the density does not converge to zero. There the thin layer does not give the substantial deposit to the impulse of swarm because in each small cube there is only one sample from thin layer. If we apply to the quantum particle the external potential $V_{pot}$, it brings the deposit to the impulse of swarm because it influence to the main part of swarm that does not belong to the thin layer. 

\section{Case of many particles}

Now we show that the method of dynamic diffusion can be generalized of the case of many quantum particles.
Let we be given the set of $n$ quantum particles, which we enumerate by natural numbers: $1,2,\ldots, n$. 
To find the efficient scheme of modelling algorithm we should apply sequentially the method of collective behaviour, at which the algorithmic reduction of quantum states is the inbuilt property. 
The most correct decoherence model, which we named absolute, says that decoherence is the reduction of quantum state 

\begin {equation}
| \Psi\rangle = \sum\limits_j\la_j|j\rangle
\label{psi}
\end{equation}
while the memory of the modelling computer cannot contain full record of this state. As we have shown above, such model gives Born's rule for calculation of probabilities by an outcome of measurement of a quantum state of system, and it proves a correctness of the given model. But such form of absolute model of decoherence cannot yet serve as heuristics for modelling algorithm as we while do not have any way of modelling of unitary quantum dynamics for many particles, except calculations within the limits of matrix algebra, and this way as we saw, is too expensive.

We will consider swarm representations of ours $n$ particles $1,2, \ldots, n $ where $S_1, S_2, \ldots, S_n $ are their swarms, corresponding to their states
 \newline 
$ | \Psi_1\rangle, | \Psi_2\rangle, \ldots, | \Psi_n\rangle$. If to consider the ensemble consisting of all these samples, it will represent a simple state of a kind 
\newline $ | \Psi_1\rangle\bigotimes | \Psi_2\rangle\bigotimes \ldots\bigotimes | \Psi_n\rangle $. But for representation of the entangled state of a kind
\begin {equation}
\Phi\rangle = \sum\limits _ {j_1, j_2, \ldots, j_n} \la _ {j_1, j_2, \ldots, j_n} |j_1, j_2, \ldots, j_n\rangle
\label{many}
\end{equation}
we need to introduce a new essential element into the method of collective behaviour. These are bonds between samples of different swarms, which we call own bonds, to distinguish them from non own bonds, which we have introduced earlier to join samples in symplexes. The basic state $j_i $ can be considered as coordinates of the particle $i $ in the corresponding configuration space. Representation of the wave function in the form (\ref {many}) means that there are bonds, which connect points $j_1, j_2, \ldots, j_n $ in one cortege. 

In the method of collective behaviour (see \cite{Oz1}) we will accept that bonds connect not spatial points, but samples of various real particles. These bonds can be written down as corteges

\begin{equation}
\bar s =(s_1,s_2,\ldots,s_n)
\label{cortege}
\end{equation}
where for any $j=1,2, \ldots, n \ s_j \in S_j $. Wave function $ | \Phi\rangle $ then is represented as set $ \bar S $ of corteges $ \bar s $ so that for any $j=1,2, \ldots,n$ and $s_j\in S_j $ there is exactly one cortege of the form (\ref {cortege}). Each cortege plays a role of the so-called world at many world interpretation of quantum theory. We consider this cortege (\ref {cortege}) as a sample of system of $n $ particles. In case of one particle we saw that action on a thin layer of the force proportional to the gradient of density of the swarm can simulate quantum dynamics. We will see that this process is directly generalised also on the case $n $ particles. We name $ \bar S $ swarm representation of the system of $n $ particles.

The swarm density $ \bar S $ is defined as 
\begin {equation}
\rho_{\bar S} (r_1, r_2, \dots, r_n) = \lim\limits_{dx\ar\infty} \frac {N_{r_1, r_2, \dots, r_n, \ dx}} {N (dx)^{3n}},
\label{density}
\end {equation}
where $ N_{r_1, r_2, \dots, r_n, \ dx} $ is a total number of the corteges which have appeared in the $3n $ dimensional cube with the side $dx $ and the centre $ r_1, r_2, \dots, r_n $, $N $ - total number of corteges.

If wave function $ | \Phi\rangle $ is the tensor product of one partial wave functions:
$$
| \Phi\rangle = \bigotimes\limits_{i=1}^{n} | \phi_i\rangle 
$$
the corresponding bonds then can be obtained by the random choice of samples from the uniform distribution $s_j\in S_j $ for everyone $j=1,2, \ldots, n $ which will thus form each cortege $s_1, s_2, \ldots, s_n $.
With such choice of corteges we will receive that the density of a corresponding swarm satisfies to Born's condition, which can for swarms be written down as   
\begin {equation}
\sum\limits_{\bar r\in D} | \langle\bar r |\Phi\rangle |^2 = \frac {N _{\bar r, \bar S}} {N}
\label {born}
\end {equation}
where $D\subset R^{3n} $, $ N_{\bar r, \bar S} $ is the total number of corteges which have appeared in area $D $. But for the entangled state $ | \Phi\rangle $ such choice of corteges for a set of swarms $ \bar S $ will not give us a condition (\ref {born}). We thus should take (\ref {born}) for definition of a choice of corteges in $ \bar S $. But for definition of a swarm we should define also speeds of all samples, namely, generalise equality (\ref {connection}) on the case of $n $ real quantum particles.  

Let $ \Psi (r_1, r_2, \ldots, r_n) $ be a wave function of system of $n $ particles, \newline
$ \Psi = |\Psi|exp (i\phi (r_1, r_2, \ldots, r_n)) $ its Euler decomposition. We will designate through $ \nabla_j\phi (r_1, r_2, \ldots, r_n) $ the gradient $ \Psi $, taken on coordinates of the particle $j $ where $j\in \{1,2, \ldots, n \} $ there is a fixed number. Generalisation of formulas (\ref {connection}) on $n $ particles looks like 
\begin{equation}
\begin{array}{ll}
&|\Psi(\bar r)|=\sqrt{\rho(\bar r)};\\
&\phi(r)=\int\limits_{\bar \g :\ \bar r_0\ar \bar r}k\bar v\cdot d\g,\\
&\bar v = a\bar\nabla\phi(\bar r),
\end{array}
\label{connection_n}
\end{equation}
where $ \bar r $ designates $r_1, r_2, \ldots, r_n $, $ \bar\nabla $ designates $ (\nabla_1, \nabla_2, \ldots, \nabla_n) $, and $ \bar \g $ is a path in $3n $ dimensional space. (\ref {connection_n}) is enough rule for the definition of the swarm on the given wave function if we agree to unite samples in corteges irrespective of their speeds. At transition from the case of one particle to the case of many particles it is necessary instead of a sample of the swarm for one particle everywhere to insert the sample of the swarm for all system of many particles. 

The external description of change of speeds will be the direct generalization of the case of one particle.
The gradient $ \bar\nabla \rho $ of the density of this swarm will cause change of speed of each sample of a separate particle in those corteges, which belong to the thin layer. Return of an impulse of the thin layer to all swarm occurs in each cell of configuration space in the same way, as in case of one particle. Dynamic smoothing of formed peaks also goes by the general rule.

Let's consider now the swarm of samples of quantum system with $n $ particles. Each of them represents Everett quantum world, and looks like $ \bar s = (s_1, s_2, \ldots, s_n) $. How such cortege receives an increment of speeds $ \Delta \bar p_{\bar s} = (\Delta p_1, \Delta p_2, \ldots, \Delta p_n) $? For this purpose it should be in the thin layer. Let it already there is. Then we will consider all other corteges lying close to $ \bar s $ in sense of the metrics of $n $ particle configuration space, and we will find the gradient of their density which is looking like: 
\begin {equation}
\bar \nabla\bar\rho = (\nabla_1, \nabla_2, \ldots, \nabla_n).
\label {nabla_}
\end {equation}

Now each sample $s_j $ receives an increment of its own impulse on $-I\nabla_j $. Cortege elements (\ref {nabla_}) are not defined by density of separate particles. They are defined by density of the swarm for all system with $n $ particles. That is influence on $ \nabla_j $ renders not only swarm density of $j $-th particle taken separately, but samples of all other particles which are connected in one cortege with an environment of $j $-th particle provided that these corteges lie close to $ \bar s $. We could consider only density of separate particles only in the case when the state is not entangled.\footnote {just for non entangled states the classical consideration of ensembles is legal.} Thus if we wish to study the general case of the entangled system in swarm representation, we should enter into consideration bonds between samples in individual swarms, in particular, for corteges belonging to a thin layer. 
As in the case of one particle, we redefine thin layer at each step.

Here again in case of a non entangled state the role will play only affinity of samples of separate particles. Bonds between the different samples entering into one cortege will be involved only if we have an entangled state. 
Thus, the thin layer in a case of the entangled state of $n $ particles also cannot be received as a simple combination something like ''thin layers'' for separate particles. It is defined through bonds in corteges, that is has essentially multipartial character.

Let $m $ be the number of samples of each separate particle. Then, owing to our condition of uniqueness of the cortege containing the given sample of any particle, we have exactly $m $ various corteges, which are not overlapping pair wise 
 If we launch the number of real particles $n $ to infinity, having left $m $ to constant, we with growth of number of real particles $n $ will receive the increasing deviation of our model from the exact solution of the corresponding Shredinger equation. Thus, fixing $m $ means presence of the internal factor of decoherence, which is independent from any ''environment''. On the other hand, complexity of model will linearly increase with growth $n $, instead of exponential, as in the standard formalism. That is the method of collective behaviour realises absolute model of decoherence which in it comes owing to limitation of memory of the modelling computer.

Acceleration of a thin layer as a result of action of a gradient of density of a swarm in case of limitation of number of samples gets the specific form. If to divide configuration space of each real particle on $s $ cubes the configuration space for $n $ particles will appear divided on $s^n $ cubes. Therefore we can speak about any ''calculation of a gradient of density'' in Bohm' sense only in that case, when $m\gg s^n $ that is when it is not a lot of real particles. If this inequality is not true, division of the force operating on a separate cortege from the swarm on ''analytically described component'' and ''processing of peaks'' loses meaning. ''Peaks'' will be actually everywhere, and it is necessary to describe process of their dynamic smoothing correctly. 

Now we take up the mechanism of changing of speeds for samples of $n$ particle system. It is formulated as in the case of one particle, but the close samples will be replaced by close corteges. Namely, a cortege we treat as a symplex of the first order. If two corteges $\bar s_1$ and $\bar s_2$ are close in the sense of $n$ particle configuration space, e.g. they belong to the same cell in this space. Then in the instant when the distance between then is minimal the own bonds between the corresponding samples are established and we obtain the $n$ particle symplex of the second order. We call the sequence of these one particle bonds an $n$ particle bond, where one particle bonds are its components. A symplex for the $n$-th order arises from a symplex of $n-1$-th order and a sample by the same procedure. The decay of all one particle components of a bond happens simultaneously, that is we can speak about the decay of $n$ particle bond in a given time instant. The decay results in the flying away a sample, etc. Thin layer consists of samples, which move from cell to cell at the short time frame. With this mechanism we again obtain the action of gradient of $n$ particle swarm of corteges, which acts on the thin layer. 

\section{Conclusion}

We show how quantum unitary dynamics can be simulated in terms of a swarm of classical point wise samples. Mechanism of speedup samples rests on the new object - bonds, which are absolutely rigid connections joining samples into symplexes. A symplex is a set of close samples, which roll with high speed, whereas a symplex as a whole moves with low speed. Quantum dynamics comes from the creation and annihilation of bonds that formed a kind of Poisson process. In course of it samples fly out from their parent symplexes and rapidly jump to the other close symplexes, giving them their impulses. This process results in the rapid transmission of impulse. The derivative of a stream of samples through the small border between cells turns proportional to the density gradient, where the coefficient depends on the grain of spatial resolution as $\d x^{-3}$. 

This dependence makes impossible to launch the grain to zero and hence, to use the exact analytic apparatus to the swarm dynamics. The reason of it lies in the nature of diffusion swarm - it consists of two fractions: slow (symplexes) and fast (separate samples, fling with the speed of light). The adventage of dynamic diffusion is in that this approach to simulation of quantum dynamics can be easily generalized to the many particle case, because the mechanism of samples speedup depends only on the density in the current point, but not on the closest vicinity of it. In this generalization a sample of the whole $n$ particle system is the cortege, consisting of samples of the separate particles. This is the idea of collective behavior for swarm quantum dynamics. These corteges do not intersect and their total number equals the total number of samples of one particle. The limitation of this number plays the role of decoherence when the number of real particles converges to infinity. In this case the approximation via collective behavior becomes farer and farer from unitary quantum dynamics. 

Rotation of samples inside of symplexes bears a resemblance to the spin of particle (see \cite{E}). We cannot claim that there is the conventional quantum mechanical spin, this spin we can call ''hydrodynamic spin''. The establishing of the relation of this ''spin'' with the usual spin represents the interesting problem. 

The check of validity of dynamic diffusion is the subject of computer simulation. The main problem arises with the area where $\rho\approx 0$. For example, for the particle in the rectangular potential hole its ground state has the form $sin^2ax$. The influence if a wall we represent as the elastic reflection of samples and symplexes from the walls. The density then will have peaks along all area, and they move the faster the less is the density. Just the existence of these peaks preserves the form of $sin^2$ for the sufficiently long time in comparison to the other functions. The other example is oscillation of a particle between two potential holes separated by high potential barrier. The huge velosity of samples in thin layer makes possible to overcome this barrier, if the general speed created by slow symplexes has the same direction. If the direction of this slow part of a swarm is reverse, thin layer cannot overcome the barrier. This effect gives the known oscillation, following from Shredinger equation. 

The further development of collective behavior lies to the effect of decoherence in complex systems (see \cite{OO}, \cite{OAO}). Here the approximate representation of quantum dynamics in the form of dynamic diffusion will have the adventage over Bohm approach, because it does not require the smooth shape of the density.

\end{document}